\documentclass[prl,twocolumn,showpacs,preprintnumbers,amsmath,amssymb]{revtex4}

\usepackage{graphicx}
\usepackage{dcolumn}
\usepackage{bm}

\begin{document}


\title{Why does wurtzite form in nanowires of III-V zinc-blende semiconductors?}

\author{Frank Glas}
\email{frank.glas@lpn.cnrs.fr} \affiliation{CNRS - Laboratoire de Photonique et
de Nanostructures, route de Nozay, 91460 Marcoussis, France}

\author{Jean-Christophe Harmand}
\affiliation{CNRS - Laboratoire de Photonique et de
Nanostructures, route de Nozay, 91460 Marcoussis, France}

\author{Gilles Patriarche}
\affiliation{CNRS - Laboratoire de Photonique et de Nanostructures, route de
Nozay, 91460 Marcoussis, France}

\begin{abstract}
We develop a nucleation-based model to explain the formation of the wurtzite
(WZ) phase during the vapor-liquid-solid growth of free-standing nanowires of
zinc-blende (ZB) semiconductors. Nucleation occurs preferentially at the edge
of the solid/liquid interface, which entails major differences between ZB and
WZ nuclei. Depending on the pertinent interface energies, WZ nucleation is
favored at high liquid supersaturation. This explains our systematic
observation of ZB during early growth.
\end{abstract}

\pacs{68.65.La,64.60.Qb,81.05.Ea,81.15.Kk,64.70.Nd} \keywords{nanowires,
nanorods, nanowhiskers, epitaxy, growth from the vapor phase,
vapor-liquid-solid, III-V semiconductors, crystal structure, wurtzite,
zinc-blende, sphalerite, nucleation, phase transition, triple line,
transmission electron microscopy}

\maketitle

Free-standing wires with diameters ranging from hundreds down to a few
nanometers are nowadays commonly fabricated from a large range of semiconductor
materials \cite{Koguchi1992,Duan2000,Persson2004,Harmand2005,Johansson2006}.
These nanowires (NWs) have remarkable physical properties and many potential
applications. The present work deals with the epitaxial growth of NWs of III-V
semiconductors on a hot substrate. Metal catalyst nanoparticles deposited on
the substrate before growth define the wire diameter. According to the
vapor-liquid-solid (VLS) growth mechanism, the atoms are fed from the vapor
phase to the solid wire through this particle (or droplet), which remains
liquid during growth \cite{Wagner1964}.

We consider III-V compounds which, under bulk form, adopt the cubic zinc-blende
(ZB) crystal structure \cite{Yeh1992} (although some non-ZB high-pressure
phases \cite{Ackland2001} may be metastable at atmospheric pressure
\cite{Mcmahon2005}), leaving aside nitrogen-based NWs. We discuss the usual
case of NWs grown on a [111]B (As-terminated) face of the ZB substrate.
Probably the most surprising feature of these NWs is that, in contrast to their
bulk counterparts, they often adopt the hexagonal wurtzite (WZ) structure. This
was observed for most ZB III-V materials and growth techniques
\cite{Koguchi1992,Persson2004,Harmand2005,Mohan2005,Soshnikov2005}. However,
although often dominantly of WZ structure, the NWs usually contain stacking
faults (SFs) and sequences of ZB structure. The coexistence of two phases is
clearly a problem for basic studies as well as applications, so that phase
purity control is one of the main challenges of III-V NW fabrication.

The surprising prevalence of the WZ structure in III-V NWs has not been
explained satisfactorily so far. Here, based on new experimental observations,
we propose an explanation of the occurrence of the WZ structure and develop a
model predicting quantitatively in which growth conditions it should form. We
consider the specific case of gold-catalyzed GaAs NWs grown by molecular beam
epitaxy (MBE) on a GaAs substrate but we expect our model and our conclusions
to remain valid for any ZB III-V compound and any growth method.

Let us start with briefly reviewing previously proposed explanations.
Calculations give the difference $\delta w$ in cohesive energy between ZB and
WZ bulk GaAs as about 24 meV per III-V pair at zero pressure \cite{Yeh1992}. It
has been argued that this favoring of the ZB form might be offset in NWs of
small diameter by the large relative contribution to the total energy of either
the lateral facets \cite{Leitsmann2006} or the vertical edges separating the
latter \cite{Akiyama2006} (provided the specific energies of these features are
less for WZ than for ZB). This naturally leads to critical radii under which WZ
NWs should be more stable than ZB NWs. For instance, Akiyama \emph{et
al.}~calculated a critical radius of 5.6 nm for GaAs NWs. These approaches have
in common to treat the energetics of \emph{fully formed} NWs and to predict
critical radii far too small to explain the occurrence of WZ in NWs with radii
up to at least 100 nm.

On the other hand, from the very beginnings of VLS studies, it has been argued
that the two-dimensional (2D) nucleation of new solid layers from the
supersaturated liquid was of paramount importance \cite{Mutaftschiev1965} and
most theories of NW growth take nucleation into account
\cite{Givargizov1975,Dubrovskii2004,Kashchiev2006}. The fact that the faults in
each phase and those separating ZB and WZ regions are perpendicular to the
growth axis, in other words that each monolayer (ML) of III-V pairs is uniform
in structure and orientation, strongly suggests that, once a nucleus of
critical size is formed, it rapidly spreads out laterally over the whole
solid/liquid (SL) interface \cite{Soshnikov2005,Johansson2006}, unless the wire
is very wide . If so, the reason for the formation of the WZ phase should not
be searched in the energetics of fully formed NWs
\cite{Leitsmann2006,Akiyama2006} but instead in the preferential formation of
WZ nuclei at the SL interface. This is the aim of the present work.

\begin{figure}[t]
\includegraphics [width=50mm]{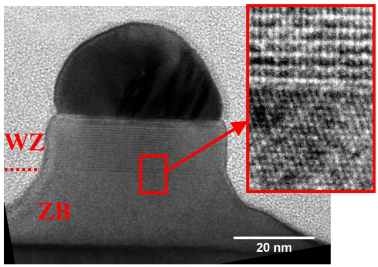}
\caption{\label{fig:TEMBeginEnd} (Color online) TEM image of a short GaAs
nanowire with high resolution close-up of the ZB$\rightarrow$WZ transition
zone.}
\end{figure}

Given the prominence of WZ in GaAs NWs, it is interesting to find two instances
where the ZB structure \emph{systematically} appears. The first one corresponds
to the initial stage of NW growth. By growing GaAs NWs for short times, we
ensured that the foot of each NW could be observed clearly before being buried
by the 2D layer which grows concomitantly albeit more slowly between the NWs.
The cross-sectional images obtained \emph{ex situ} by transmission electron
microscopy (TEM) prove that the whole NW is initially pure ZB. Growth then
switches abruptly to WZ stacking (Fig.~\ref{fig:TEMBeginEnd}). Scanning
electron microscopy shows that, at this early stage, the NWs are pyramids with
triangular bases and tilted lateral facets. These become vertical at the
ZB$\rightarrow$WZ transition (Fig.~\ref{fig:TEMBeginEnd}).

The second observation is that when we terminate MBE growth by switching off
the Ga flux while maintaining an As flux, a section of NW grows that
systematically adopts the ZB structure. This effect, already reported for GaAs
NWs grown by chemical beam epitaxy \cite{Persson2004}, is interpreted as a
partial consumption of the Ga dissolved in the gold particle to form the
terminal section of the NW.

These two situations where ZB forms have in common to be transient growth
phases during which the supersaturation of Ga (and possibly As) in the liquid
is less than during steady NW growth. Before growth, the deposited Au droplets
dissolve the substrate locally to achieve equilibrium with it: the
supersaturation is zero. When vapor fluxes are turned on, the supersaturation
increases until a permanent regime settles. Symmetrically, during growth
termination, the Ga concentration in the droplet, and hence supersaturation,
decrease, since the atoms used to build the NW are not replaced. This strongly
suggests that ZB systematically forms when the supersaturation is less than
some critical value and, conversely, that WZ formation requires a \emph{high
supersaturation}.

This confirms the importance of nucleation. Indeed, according to nucleation
theory, the work needed to form solid nuclei from a fluid phase is maximum for
a critical nucleus size \cite{Kashchiev2000}. If this energy barrier is less
for a certain crystal structure than for another, the first may nucleate
preferentially even if the second one is more stable in bulk form. Since
critical sizes and energy barriers scale inversely with supersaturation, our
observations point to a preferential formation of WZ when the critical nuclei
are small. This recalls the preference for WZ of NWs of small radius
\cite{Akiyama2006,Leitsmann2006} but we cannot simply assume it to hold for
nuclei. Instead, we should compare the probabilities of forming various nuclei
from the liquid phase. As a first approximation, we shall compare their
formation energies, which largely determine these probabilities. In order not
to obscure our demonstration, we keep as far as possible to continuous
nucleation theory (CNT) \cite{Kashchiev2000}, resorting to an atomistic picture
only when necessary. We proceed in two stages. We first show that nucleation
must occur at the edge of the SL interface (the triple solid/liquid/vapor line)
rather than elsewhere in the SL interface. We then show that, along this line,
the formation of WZ nuclei may be favored over that of ZB nuclei.

\begin{figure}[t]
\includegraphics [width=60mm]{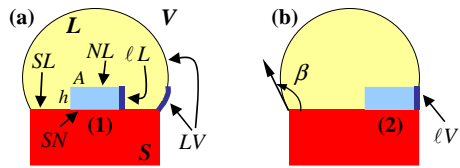}
\caption{\label{fig:TripleLine}  (Color online) (a) Nucleus at the NW/liquid
interface, with interfaces of interest. (b) Transferring the nucleus to the
triple line eliminates and creates interfacial areas (thick lines).}
\end{figure}

Let us consider the interface between a NW (hereafter called substrate) and the
liquid droplet. According to CNT, forming there a solid epitaxial 2D island of
height $h$ (1 ML), perimeter length $P$ and upper area $A$
(Fig.~\ref{fig:TripleLine}(a)) from the liquid phase, involves a change of free
enthalpy:
\begin{eqnarray}
\Delta G = - A h \Delta \mu  + P h \gamma_{\ell L} + A \left( \gamma_{NL} -
\gamma_{SL} + \gamma_{SN} \right) \label{eq:deltaGstandard}
\end{eqnarray}
In Eq.~(\ref{eq:deltaGstandard}), $\Delta \mu > 0$ is the difference of
chemical potential for III-V pairs between liquid and solid phases, per unit
volume of nucleus; $\gamma_{\ell L}$ is the energy per unit area of the
\emph{lateral} interface between nucleus and liquid; $\gamma_{SL}$,
$\gamma_{SN}$ and $\gamma_{NL}$ are, respectively, the energies per unit area
of the substrate/liquid, substrate/nucleus (SN) and upper nucleus/liquid (NL)
interfaces (Fig.~\ref{fig:TripleLine}(a)).

\begin{figure}[b]
\includegraphics [width=55mm]{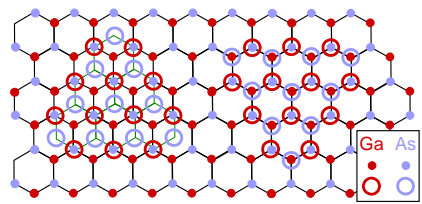}
\caption{\label{fig:NucleiZBWZ} (Color online) A given nucleus (open symbols)
in ZB (left) and WZ (right) positions on top of a (111)B ML (disks).}
\end{figure}

A given nucleus (set of atoms with fixed relative positions) of ML height
cannot be said to be of ZB or WZ structure. It is only the \emph{orientational
positioning} of the nucleus with respect to the previous ML which determines if
the stack of 2 MLs formed by adding the nucleus is of the type found in ZB or
WZ crystals (Fig.~\ref{fig:NucleiZBWZ}). In the former case (hereafter 'ZB
position') the GaAs$_4$ tetrahedra have the same orientation if the Ga atom
belongs either to nucleus or to previous ML whereas tetrahedra and nucleus are
rotated by an odd multiple of $\pi / 3$ in the latter case ('WZ position')
\cite{Yeh1992}. ZB and WZ \emph{sequences} require the nucleation of
\emph{each} ML in, respectively, ZB and WZ position \emph{with respect to the
previous ML}.

Whatever the position of the nucleus, interfaces SL and NL have the same atomic
configuration (Fig.~\ref{fig:NucleiZBWZ}) so that $\gamma_{SL} = \gamma_{NL}$.
On the other hand, nucleation in WZ position costs some SN interfacial energy
($\gamma_{SN}^{WZ} = \gamma_{F} > 0$) whereas nucleation in ZB position does
not ($\gamma_{SN}^{ZB} = 0$). Since forming a WZ sequence on a ZB substrate
creates a SF, $\gamma_{F}$ is sometimes taken \cite{Johansson2006} as half the
SF energy in the bulk ZB phase \cite{Takeuchi1999}. Finally, from
Eq.~(\ref{eq:deltaGstandard}), the formation enthalpies of a given nucleus in
ZB or WZ position are $\Delta G^{ZB,WZ} = - A h \Delta \mu + P h \gamma_{\ell
L}+ A \gamma_{SN} ^{ZB,WZ}$. Since $\Delta G^{WZ} - \Delta G^{ZB} = A
\gamma_{F} > 0$, ZB nucleation is favored.

In the foregoing discussion, we did not specify if $\gamma_{\ell L}$ refers to
ZB or WZ because the nucleus, which clearly has the same edges in both
positions, was assumed to be laterally surrounded by the liquid. However, we
now show that nucleation should take place at the triple line. Let us compare
the formation of a given nucleus at two different locations
(Fig.~\ref{fig:TripleLine}): its lateral surface is either entirely surrounded
by the liquid (1) or partly surrounded by the vapor because of nucleation at
the triple line (2). We now have to distinguish the specific energies
$\gamma_{\ell L}$ and $\gamma_{\ell V}$ of the lateral nucleus/liquid and
nucleus/vapor interfaces. The key point is that shifting the nucleus from
location (1) to location (2) \emph{at constant liquid volume} has a major
effect: it eliminates part of the liquid/vapor interface and replaces it by
nucleus/vapor interface (Fig.~\ref{fig:TripleLine}). Suppose that forming an
area $s$ of nucleus/vapor interface eliminates an area $\tau s$ of liquid/vapor
interface, of energy $\gamma_{LV}$. If $\alpha$ is the fraction of the island
perimeter in contact with the vapor, the formation enthalpy of the nucleus now
is:
\begin{eqnarray}
\Delta G & = & - A h \Delta \mu  + P h \left[ \left( 1 - \alpha \right)
\gamma_{\ell L} + \alpha \left( \gamma_{\ell V} - \tau \gamma_{LV} \right)
\right] \nonumber \\  & & + A \gamma_{SN} \label{eq:deltaGLV}
\end{eqnarray}
where, as above, $\gamma_{SN} = 0 \textrm{ or } \gamma_{F}$ for, respectively,
ZB or WZ nuclei. For a given nucleus, the difference in formation enthalpies
between locations (1) and (2) is $\Delta G_2 - \Delta G_1 = \alpha P h \left(
\gamma_{\ell V} - \gamma_{\ell L} - \tau \gamma_{LV} \right)$. Factor $\tau$
cannot be calculated exactly. We estimate it by considering an artificially
axisymmetric nucleus with a vertical lateral surface contacting the vapor along
the whole triple line. This geometry preserves a spherical liquid/vapor
interface. It readily yields $\tau = \sin \beta$, with $\beta$ the contact
angle between droplet and substrate (Fig.~\ref{fig:TripleLine}). In all our
samples, $90^{\circ} \le \beta  \le 130^{\circ}$ (after growth) so that $0.85
\le \tau \le 1$.

Hence, a given nucleus tends to form at the triple line if $\gamma_{\ell V} -
\gamma_{\ell L} - \tau \gamma_{LV} < 0$, with $\tau \simeq 1$. Before growth,
the contact angles between our liquid droplets and bulk GaAs are close to $\pi
/ 2$, which implies (from Young's equation) that the solid/liquid and
solid/vapor interface energies are close for (111)B surfaces. Assuming that
this also holds for the lateral nucleus faces yields $\gamma_{\ell L} \simeq
\gamma_{\ell V}$. In turn, $\gamma_{LV}$ should lie between the surface
energies of pure liquid Au and Ga (1.14 and 0.72 $\textrm{J.m}^{-2}$
\cite{Zangwill1988}). Hence, the above inequality is safely satisfied and the
critical nuclei should form at the edge of the droplet. In short, forming the
nucleus there is advantageous because it eliminates a portion of the
\emph{pre-existing} droplet surface; this largely outweighs the replacement of
part of the lateral nucleus/liquid interface by a possibly slightly costlier
nucleus/vapor interface. Note that nucleation at the triple line in GaP NWs has
previously been argued for on an entirely different basis, namely the low
solubility of phosphorus in gold \cite{Johansson2006}. Our argument is of
general validity and would even apply to solid catalyst particles
\cite{Persson2004}.

\begin{figure}[b]
\includegraphics [width=80mm]{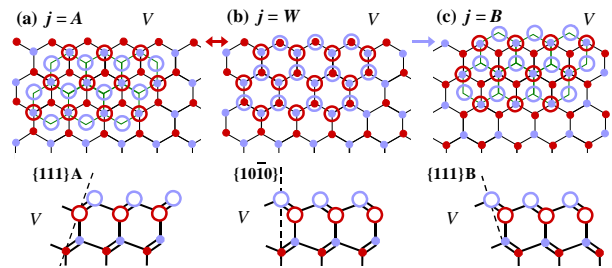}
\caption{\label{fig:Bords} (Color online) Top (top row) and side (bottom row)
views of a given nucleus in ZB (a) and WZ (b) positions at a Ga edge and in ZB
position at an As edge (c). Edges are indicated by arrows (top), the vapor side
by $V$ and non-vertical dangling bonds by segments (bottom). Key as in
Fig.~\ref{fig:NucleiZBWZ}.}
\end{figure}

To demonstrate the advantage of WZ over ZB nucleation \emph{at the triple
line}, we consider the atomic structure of the edges of the top facet of the NW
and of the nucleus. We simply select the low energy configurations discussed
below by restricting ourselves to at most one non-vertical dangling bond per
atom and to stoichiometric nuclei. At the first ZB$\rightarrow$WZ transition
(Fig.~\ref{fig:TEMBeginEnd}), given the pyramidal shape of the NW foot, the
three edges of the top facet must be rows of Ga atoms. A given nucleus can
again be placed there in ZB or WZ position, but this now entails major
differences. The ZB nucleus must itself have a Ga edge at the triple line, so
that the edges of the top NW facet and of the nucleus develop a tilted
$\{111\}$A-type lateral facet (Fig.~\ref{fig:Bords}(a)). The WZ nucleus has an
As edge at the triple line; the lateral facet is then vertical and of
$\{10\bar{1}0\}$ WZ type (Fig.~\ref{fig:Bords}(b)). Since WZ \emph{sequences}
require the repeated nucleation of each ML in WZ position, the latter must
therefore be favored at both Ga and As edges. This is also needed when WZ NWs
with six $\{10\bar{1}0\}$-type vertical facets have started developing since
their top facets have three edges of each type. However, at an As edge, ZB and
WZ positions also differ: in ZB position, the nucleus itself has an As edge and
generates a $\{111\}$B-type tilted lateral facet (Fig.~\ref{fig:Bords}(c)),
whereas in WZ position (not shown) it has a Ga edge and the lateral facet is
again vertical, of $\{10\bar{1}0\}$ WZ type.

Let us first compute the energy changes $\Delta_j$ upon transferring a given
nucleus from location (1) (Fig.~\ref{fig:TripleLine}(a)) to three possible
locations at the triple line (Fig.~\ref{fig:Bords}), one in WZ position with a
lateral $\{10\bar{1}0\}$ facet ($j = W$) and two in ZB position with lateral
$\{111\}$A ($j = A$) or $\{111\}$B ($j = B$) facets. When the lateral facets
are tilted toward ($j = A$) or away from ($j = B$) the NW axis, the transfers
also modify the areas of the solid/liquid and liquid/vapor interfaces, at fixed
liquid volume. This can again be estimated by first considering toroidal nuclei
preserving spherical liquid/vapor and axisymmetric SL interfaces, and then
taking $\Delta_j$ as proportional to the fraction $\alpha P$ of the perimeter
of the actual nucleus in contact with the vapor. We find $\Delta_j = \alpha P h
\left( \widetilde{\gamma}_j - \gamma_{\ell L} - \gamma_{LV} \sin \beta \right)$
where $\widetilde{\gamma}_j = \gamma_j / \cos \theta_j + \left( \gamma_{LS} +
\gamma_{LV} \cos \beta \right) \sin \theta_j$, with $\gamma_j$ the interface
energy between vapor and lateral facet and $\theta_j$ the angle of the latter
with the vertical ($\theta_W = 0, \theta_B = - \theta_A = 19.5 ^\circ$). Our
$\widetilde{\gamma}_j$ have the same expression as the energies calculated by
Ross \emph{et al.}~in a 2D model of facetted NWs \cite{Ross2005}, but they
pertain to ML high nuclei.

Although it is not essential, we now specify that the nuclei are equilateral
triangles of side $D$, one of them at the triple line ($\alpha = 1 / 3$). Their
formation enthalpies are:
\begin{eqnarray}
\Delta G_j & = & - \frac{\sqrt{3}}{2} D^2 h \Delta \mu  + 3 D  h \gamma_{\ell
L} - D h \left( \gamma_{\ell L} + \gamma_{LV} \sin \beta \right) \nonumber \\
& & + D h \widetilde{\gamma}_j + \frac{\sqrt{3}}{2}D^2 h \gamma_{SN}
\label{eq:deltaGEquilateral}
\end{eqnarray}
Only the last two terms differ between nuclei and the last one is non-zero only
in WZ position. Maximizing $\Delta G_j$ with respect to $D$ yields the critical
size $D_j^{\star}$ and the energy barriers $\Delta G_j^{\star} = \Delta G_j
\left( D_j^{\star} \right)$ for each nucleus:
\begin{eqnarray}
\Delta G_W^{\star}  = \frac{\sqrt{3}}{6} \frac{\Gamma_W^2}{\Delta \mu -
\frac{\gamma_F}{h}} \ , \ \Delta G_k^{\star}  = \frac{\sqrt{3}}{6}
\frac{\Gamma_k^2}{\Delta \mu} \textrm{ for } k=A,B \label{eq:CriticalDeltaG}
\end{eqnarray}
where $\Gamma_j = 2 \gamma_{\ell L} + \widetilde{\gamma}_j - \gamma_{LV} \sin
\beta$ is an effective edge energy. WZ nuclei dominate if $\Delta G_W^{\star} <
\Delta G_k^{\star}$ for $k=A,B$. This requires two conditions. The first one,
$\widetilde{\gamma}_W < \widetilde{\gamma}_k$ for $k=A,B$, is material-related
and involves only interface energies. A second, growth-related, condition is
that the supersaturation be larger than a critical value, $\Delta \mu^{\star} =
\max_{k=A,B} \left( \frac{\Gamma_k^2}{\Gamma_k^2 -
\Gamma_W^2}\frac{\gamma_F}{h} \right)$, to overcome the SF.

As a first approximation, we estimate the ZB $\widetilde{\gamma}_j$ energies
from those of (111)A and B surfaces computed for As-rich (MBE) vapors, namely
0.82 and 0.69 $\textrm{J.m}^{-2}$ \cite{Moll1996}. In the extreme cases of drop
surfaces of pure Ga (surface segregation of the low energy atom) and pure Au,
this gives respectively $\widetilde{\gamma}_A = 0.76 $ and $0.83 \textrm{
J.m}^{-2}$ and $\widetilde{\gamma}_B = 0.84 $ and $0.77 \textrm{ J.m}^{-2}$,
well above the low $\gamma_{111B}$ energy. Conversely, $\widetilde{\gamma}_W =
\gamma_{10\bar{1}0}$ is unknown for As-rich vapors. According to the previous
discussion, WZ forms because $\widetilde{\gamma}_W < \widetilde{\gamma}_A
\textrm{ and } \widetilde{\gamma}_B$. This hypothesis is strengthened by
calculating the critical supersaturations for a plausible range of such values,
$0.7 \le \widetilde{\gamma}_W \le 0.75 \textrm{ J.m}^{-2}$. For \emph{e.g.~}a
Ga drop surface and $\beta = 120^{\circ}$, $\Delta \mu^{\star}$ ranges between
230 and 1570 meV, which is indeed of the order of our experimental
supersaturations (several 100 meV).

In summary, we developed a nucleation-based model to explain the occurrence of
the WZ phase in nanowires of ZB semiconductors, at least at certain stages of
growth. A key and general result is that 2D nucleation takes place
preferentially at the edge of the solid/liquid interface. When formed at this
triple line, WZ and ZB nuclei present major differences and WZ nucleation is
actually favored for certain ranges of the interface energies involved. In
addition, the supersaturation of the liquid must be high enough, in agreement
with our experimental results. Our aim was to identify important effects and
parameters, not yet to give a complete description of the complex interplay of
the two phases. We now intend to calculate the actual nucleation probabilities
(including the effects of temperature and geometry), evaluate more precisely
the energies of various nuclei (including non-stoichiometric ones) forming on
NWs with different cross-sections, and take into account growth conditions in
more details, in particular the supersaturation of each atomic species which
appears here only indirectly via surface energies.

\begin{acknowledgments}
This work was partly supported by the SANDIE Network of Excellence of the
European Commission (Contract No. NMP4-CT-2004-500101).
\end{acknowledgments}

\end{document}